\documentclass[letterpaper, 10 pt, conference]{ieeeconf}  

\IEEEoverridecommandlockouts                              

\usepackage{amssymb,amsmath} 
\usepackage{amsthm} 
\usepackage{graphicx}
\usepackage{subfig}
\usepackage{algorithm}
\usepackage{algorithmic}
\usepackage{enumerate}
\usepackage{cite}
\usepackage[table]{xcolor}
\usepackage{thmtools}

\usepackage{color}

\newcommand{\calP} {\mathcal{P}}

\newcommand{\calF} {\mathcal{F}}
\newcommand{\calG} {\mathcal{G}}

\newcommand{\calU} {\mathcal{U}}
\newcommand{\rmProb}{\mathrm{Prob}}

\newcommand{\argmax}[1]{\underset{#1}{\arg\max}~}

\renewcommand\thmcontinues[1]{cont.}

\newtheorem{problem}{Problem}

\theoremstyle{definition}
\newtheorem{example}{Example}

\newcommand{\graycell}{\cellcolor{gray!25}}

\title{\LARGE \bf
	Control with Probabilistic Signal Temporal Logic
}

\author{Chanyeol Yoo and Calin Belta 
\thanks{The authors are with the Department of Mechanical Engineering, Boston University, Boston, MA 02215, USA {\tt \{chanyeol,cbelta\}@bu.edu}.}
\thanks{This work was partially supported by the NSF under grants NRI-1426907 and CMMI-1400167.}%
\thanks{This work is currently under review in the 2016 Americal Control Conference (ACC 2016, submitted on September, 30, 2015).}
}

\begin{document}

\maketitle
\thispagestyle{empty}
\pagestyle{empty}

\begin{abstract}


Autonomous agents often operate in uncertain environments where their decisions are made based on beliefs over states of targets. We are interested in controller synthesis for complex tasks defined over belief spaces. Designing such controllers is challenging due to computational complexity and the lack of expressivity of existing specification languages. In this paper, we propose a probabilistic extension to signal temporal logic (STL) that expresses tasks over continuous belief spaces. We present an efficient synthesis algorithm to find a control input that maximises the probability of satisfying a given task. We validate our algorithm through simulations of an unmanned aerial vehicle deployed for surveillance and search missions.

\end{abstract}

\section{INTRODUCTION}

In recent years, there has been an increased interest in using formal methods in robot motion planning and control \cite{belta_ras_2007, ras_2011, kavraki_ras_2011, wongpiromsarn2009receding}. Temporal logics, such as Linear Temporal Logic (LTL), Computation Tree Logic (CTL), and their probabilistic versions \cite{Baier2008, baier1998algorithmic} have been shown to be expressive enough to capture a large spectrum of robotic missions. Model checking and synthesis algorithms have been successfully to generate motion plans and control strategies from such specifications. Most of the current works, however, do not capture the uncertainty that is inherent in real-world applications. Autonomous agents usually operate in uncertain environments with limited and possibly corrupted information.

In this paper, we propose a specification language called \emph{probabilistic Signal Temporal Logic} (PrSTL), which is a probabilistic extension of an existing temporal logic, called Signal Temporal Logic (STL) \cite{maler2004monitoring}. The specifications are interpreted over a \emph{belief space} (the space of probability distributions over states of an environment)~\cite{kaelbling2013integrated, nguyen2013energy} about the locations of a set of targets. We propose a receding horizon control strategy that maximizes the probability of satisfying the specification. The procedure involves iterations consisting of observations and belief updates using Bayes' rule.
We include illustrative simulation examples involving surveillance and search and rescue.

{\bf Contribution and Related Work}
Temporal logics have been used widely in robotic task planning~\cite{belta_ras_2007,ras_2011,belta_ras_2011,Kress-Gazit2009}. One of the most popular logic is \emph{linear temporal logic} (LTL), which provides an expressive mean to specify complex robotic tasks such as converge, sequencing, conditions and avoidance~\cite{Kress-Gazit2009}. These tasks can be combined further to form more complex tasks. Given an LTL specification, a provably-correct controller can be automatically generated for finite systems by using existing synthesis algorithms~\cite{Baier2008}. Examples of robotics applications include ~\cite{Smith2011,  yoo2015online, Ding2011, Kloetzer2007, T2013, Ulusoy2013, svorenova2013optimal}. These works consider static environments and assume that the robots have full knowledge over the environment. Hence, the controllers have to be re-synthesised from scratch if changes are made. Also, the correctness property may be violated if the knowledge over the environment is not accurate.

In order to operate robots in dynamic environments, a fragment of LTL called \emph{generalized reactivity} (GR(1)) has been used to synthesise reactive controllers~\cite{Kress-Gazit2009, Piterman2006, Wolff2013}. This game-theoretic approach considers non-deterministic changes in the environment and guarantees the satisfaction under all allowed environment changes. However, since all robot and environment behaviours are symbolically encoded as part of an LTL formula, any unexpected changes in dynamics cannot be captured during execution and complex system dynamics cannot be described accurately.
This approach also assumes that the robots have full knowledge of the environment at all times but the assumption is easily violated in practice.
Furthermore, this method only considers the worst case. Hence, the solution is often conservative and may not find a solution at all for practical cases.

Instead of modelling the behaviour of environments by non-determinism, probabilistic uncertainty is also considered in robotic task planning problems. Two popular temporal logic forms for this purpose are \emph{probabilistic computation tree logic} (PCTL)~\cite{Chanyeol2012, Chanyeol2013, Lahijanian2012, svorenova2013optimal} and \emph{probabilistic linear temporal logic} (pLTL)~\cite{Ding2011}.
These logics are capable of expressing tasks for systems with probabilistic transitions while assuming that the transition results are precisely known and that the regions of interest are static and known in advance.
Their semantics is defined over Markov decision processes (MDPs). The objective is to find a control policy that maximises the probability of satisfying a given specification.

Extended work considers non-static environments where the behaviour of adversarial environmental states are also modelled by MDPs~\cite{Wolff2013}. Even further, in~\cite{MedinaAyala2011}, environmental states are modelled by \emph{mixed observable} MDPs where some internal transitions are not visible from the outside. The solution to this problem considers the beliefs on the internal states of the environment. \emph{Partially observable} MDPs (POMDPs) allows for more general abstraction of hidden internal transitions.
Even though recent results show that LTL control synthesis over control strategies with finite memory is decidable \cite{chatterjee2013decidable}, the solutions are expensive and do not capture environmental changes. 

\emph{Signal temporal logic} (STL) is a time-bounded temporal logic developed monitor systems with continuous dynamics. Its semantics are defined over continuously valued signals~\cite{maler2004monitoring}. A satisfaction of an STL specification is determined by evaluating its \emph{degree of robustness}: a measure of how well a signal satisfies the given specification. To the best of our knowledge, no probabilistic forms of STL exist. In order to specify tasks over an uncertain environment using STL and quantitatively evaluate a signal with respect to a specification, one possible approach is to compute the \emph{expected degree of robustness} (i.e., average robustness over an infinite set of signals). However, since computing robustness for temporal operators requires $\min$ and~$\max$ operations, analytical evaluation of the expected robustness is not trivial. Also, using such an approach would require the environment to be deterministic while transitions are assumed to be probabilistic.
Our interest in this paper is in estimating true states of targets in an environment. Since the estimates over the target states are given in the form of probability distributions, it makes more sense to evaluate signals in terms of probability, not robustness.
We propose an extension to STL that can express tasks over beliefs of targets in an environment. We then use this extension to evaluate a \emph{probabilistic degree of satisfaction}.

One of the main challenges in task planning with temporal logic under uncertainty is the computational complexity of solving the synthesis problem. The time complexity of finding an optimal solution for a pLTL specification is doubly exponential in the number of propositions~\cite{Ding2011}. 
Synthesis of an STL formula also suffers from complexity blowup. Recent work in synthesis from STL specifications suggests using \emph{mixed integer linear program} (MILP)~\cite{raman2014model} and \emph{receding horizon control} (RHC) methods. 
However, since MILP is NP-hard, such algorithms are not scalable when the size of the problem is large (i.e., the number of constraints, length of formula, and time horizon).

{\bf Organisation}
The remainder of the paper is organised as follows. Sec.~\ref{sec:problem_formulation} presents the problem statement and defines the system and sensor models.
In Sec.~\ref{sec:prstl}, we define the proposed probabilistic signal temporal logic (PrSTL). In Sec.~\ref{sec:synthesis}, we present an efficient synthesis algorithm from a PrSTL formula. We then discuss the complexity of our approach in Sec.~\ref{sec:analysis} and present simulation examples in Sec.~\ref{sec:examples}. We conclude in Sec.~\ref{sec:conclusion}.


\section{PROBLEM FORMULATION} \label{sec:problem_formulation}

We consider a discrete-time dynamic agent of the form
\begin{equation} \begin{split} \label{eqn:robot_model}
	x_{t+1} = f(x_t, u_t)
	,
\end{split} \end{equation}
where $x_t \in \mathbb{R}^n$ is the continuous-valued $n$-dimensional state of the agent at discrete time~$t$, $u_t \in \mathbf{U}$ is the control input from a finite set~$\mathbf{U}$ at discrete time~$t$. We assume that the sampling time~$\Delta t$ is $1$ (i.e., $t \in \{0, 1, 2, \cdots\}$) and the state of the agent is always fully known. 
We define a \emph{run}~$\mathbf{x}$ as a sequence of agent states at time~$t$ where the prefix~$\mathbf{X}_t$ is a sequence of past states~$x_k$ (i.e., \emph{past run}) and the suffix~$\bar{\mathbf{X}}$ is a sequence of future states~$\bar{x}_k$ (i.e., \emph{future run}) given a control sequence~$\mathbf{u}$. At time~$t$, we have $\mathbf{x} = \{ \mathbf{X}_t, \bar{\mathbf{X}} \}$ given~$\mathbf{u}$, where~$\mathbf{X}_t = \{ x_0, x_1, \cdots, x_t \}$, $\bar{\mathbf{X}} = \mathbf{f}( x_t, \mathbf{u} ) = \{ \bar{x}_{t+1}, \bar{x}_{t+2}, \cdots, \bar{x}_{t+|\mathbf{u}|} \}$. Note that~$\mathbf{X}_0 = \{ x_0 \}$.

The agent is assigned a complex task~$\psi$ associated with a set of targets. The state of target~$i$ is in the form
\begin{equation} \begin{split} \label{eqn:target_model}
	\hat{x}^i_{t+1} = g^i(\hat{x}^i_t, \hat{u}^i_t)
	,
\end{split} \end{equation}
where the true $n_i$-dimensional state of the target~$\hat{x}^i_t \in \mathbb{R}^{n_i}$ evolves over time, $\hat{u}^i_t \in \mathbf{U}_i$ is a hidden control input from a finite set~$\mathbf{U}_i$, $i \in \{ 1, 2, \cdots, I \}$ and the number of targets~$I$ is finite and known in advance. We assume that the state of the agent is independent of the states of the targets. We also assume that the agent knows the model~$g^i(\cdot)$ but the exact states of the targets are not precisely known. 
Instead, the agent maintains a \emph{belief} of each target at time~$t$ defined as~$b^i_t \triangleq \mathbb{P}(\hat{x}^i_t \mid \mathbf{Z}^i_t)$, where~$\mathbf{Z}^i_t = \{z^i_0, z^i_1, \cdots, z^i_t\}$ is the history of observations made by sensors on the agent and~$z^i_t \in \{ 0, 1 \}$ (i.e., $1$ if the target~$i$ is \emph{observed} at time~$t$ and $0$ otherwise).

The belief is updated using a \emph{state estimator} when an observation is made. Assuming that observation~$z^i_t$ is independent of the history~$\mathbf{Z}^i_{t-1}$ given the true state of a target~$\hat{x}^i_{t}$ (i.e., $\mathbb{P}(z^i_t, \mathbf{Z}^i_{t-1} \mid \hat{x}^i_t) = \mathbb{P}(z^i_t \mid \hat{x}^i_t) \cdot \mathbb{P}(\mathbf{Z}^i_{t-1} \mid \hat{x}^i_t)$), the belief can be updated using Bayes' rule:
\begin{equation} \begin{split} \label{eqn:state_estimate}
	\mathbb{P}(\hat{x}^i_{t+1} \mid \mathbf{Z}^i_{t+1})	= \alpha~\mathbb{P}(z^i_{t+1} \mid \hat{x}^i_{t+1})~\mathbb{P}(\hat{x}^i_t \mid \mathbf{Z}^i_{t})
	,
\end{split} \end{equation}
where $\alpha$ is a normalising constant. The function~$\mathbb{P}(z^i_t \mid \hat{x}^i_{t})$ is the \emph{detection likelihood} which is obtained from a sensor model. 
Assuming conditional independence where observations are independent of each other given the current state, only the current observation is required to update the belief. The no detection likelihood is the complement of the detection likelihood (i.e., $\mathbb{P}(\bar{z}^i_t \mid \hat{x}^i_{t}) = 1 - \mathbb{P}(z^i_t \mid \hat{x}^i_{t})$).

The task~$\psi$ assigned to the agent is specified using a time-bounded temporal logic over a set of real-valued target beliefs. An example of such a specification is~\emph{the agent has to find two targets in 20 time steps while avoiding an obstacle. Once all the targets are found, the agent has to come back to base in 10 time steps. The probability of finding each target has to be greater than $50\%$ at all time}. This logic allows for computing the probability of satisfaction given a sequence of agent states over the target beliefs. 
In Sec.~\ref{sec:prstl}, we define such a specification language formally.

In this paper, we address the following controller synthesis problem over a belief space.
\begin{problem}[Receding horizon feedback controller synthesis over belief space]
	\label{problem:finite}
	Given a specification~$\psi$ over a finite time horizon~$H$, a past run~$\mathbf{X}_t$, a system of the form in~(\ref{eqn:robot_model}), targets of the form in~(\ref{eqn:target_model}), a state estimation model of the form in (\ref{eqn:state_estimate}) and beliefs over the targets~$B_t = \{ b^i_t \mid i = 1, 2, \cdots, I \}$, compute
	\begin{equation} \begin{split}
	\mathbf{u}^{H-t}_{t} = \argmax{\mathbf{u} \in \mathbf{U}^{H-t}} \rmProb( \{ \mathbf{X}_t, \mathbf{f}(x_t, \mathbf{u}) \}, \psi, 0 )
	,
	\end{split} \end{equation}
	where~$\mathbf{u}^{H-t}_{t} = \{ u_t, u_{t+1}, \cdots, u_{H-t} \}$ is a finite sequence of control inputs, $\rmProb$ is a function that returns the probability of satisfying a specification~$\psi$ at time~$t = 0$ given a run~$\{ \mathbf{X}_t, \mathbf{f}(x_t, \mathbf{u}) \}$.
\end{problem}

The problem is solved using a \emph{receding horizon control} (RHC) framework. RHC is an iterative control technique to solve optimisation problems in which an optimal control input over a fixed finite time horizon is determined at each time step~\cite{aksaray2015distributed, wongpiromsarn2009receding, Jones2013, Ding2010, raman2014model}. 
At each time~$t$, we compute a sequence of control inputs that maximises an objective function over a finite horizon~$H$ (i.e., between $t$ and $t+H$) and the first control input is chosen and executed. We repeat this process until the mission is complete. 
We use RHC to reduce the synthesis complexity and to rapidly react to changes in belief space.


\subsection{Examples}
We present examples using an unmanned aerial vehicle (UAV). Consider a UAV operating at a constant altitude and airspeed~$v_a$ that can be described by an equation of the form~(\ref{eqn:robot_model})
\begin{equation} \begin{split}
	\mathrm{x}_{t+1} = 
	\begin{bmatrix}
		x_{t+1}	\\	y_{t+1}	\\	\theta_{t+1}
	\end{bmatrix}
	=
	\begin{bmatrix}
		v_a \cos \theta_t	\\
		v_a \sin \theta_t	\\
		u
	\end{bmatrix}
	+ \mathrm{x}_t
	,
\end{split} \end{equation}
where~$\theta_t$ is the heading angle at time~$t$ (minutes) in an absolute Cartesian space and~$u$ is a control input from a finite set~$\mathbf{U}$. The UAV is equipped with a noisy forward-facing camera that can detect the presence of a target within the viewing range without any distance or heading information. A target is said to be \emph{in the view} when it is within the effective measuring distance~($20m$) and the angle of view ($60 \deg$). The detection likelihood of the camera~(i.e., the probability of detecting a target~$i$ at time~$t$ given a system state~$\mathrm{x}_t$ and a true state of the target~$\hat{\mathrm{x}}^i_t$) is
\begin{equation} \begin{split}
	\mathbb{P}(z^i_t \mid \hat{\mathrm{x}}^i_t, \mathrm{x}_t) = 
	\begin{cases}
		\alpha \cdot \exp(-\frac{\|\mathrm{x}_t - \hat{\mathrm{x}}^i_t\|^2}{\lambda})	&	\text{if in the view,}	\\
		0	&	\text{otherwise},
	\end{cases}
\end{split} \end{equation}
where $\alpha$ and~$\lambda$ are parameters of the camera. Note that~$\mathbb{P}(z^i_t \mid \hat{\mathrm{x}}^i_t, \mathrm{x}_t) = \mathbb{P}(z^i_t \mid \hat{\mathrm{x}}^i_t)$ when~$\mathbb{P}(\mathrm{x}_t \mid \hat{\mathrm{x}}^i_t) = \mathbb{P}(\mathrm{x}_t)$ (i.e., the agent state is independent of all the target states).

\begin{example} [Surveillance] \label{exa:surveillance}
	The UAV is required to survey three hidden targets~$\hat{\mathbf{x}}^{i}_t$ where~$i = \{ 1, 2, 3 \}$. The task is to repeatedly find each of the targets over a certain horizon. The UAV is initially given a probabilistic estimate of the targets (i.e., where they are), and the estimate is updated according to the dynamic model of the targets.
	\qed
\end{example}

\begin{example} [Prioritised search] \label{exa:search}
	
	
	The UAV is deployed in a search mission to find two suspects~$\hat{\mathbf{x}}^{i}_t$ where~$i = \{ Tom, Jerry \}$ hiding in mountains. Based on geographic data, etc, local police has computed rough probabilistic estimates of where the suspects would be. 
	Tom is given higher capture priority, hence the UAV is commanded to find Tom first and then to find Jerry.
	
	\qed
\end{example}

\section{PROBABILISTIC SIGNAL TEMPORAL LOGIC (PrSTL)} \label{sec:prstl}
In this section, we propose a probabilistic extension of signal temporal logic (STL)~\cite{maler2004monitoring} called \emph{probabilistic signal temporal logic (PrSTL)}. 
PrSTL is defined with respect to a discrete-time continuous-valued signal~$\mathbf{x}$ (i.e., a run). 
For any sequence~$\mathbf{s}$, $\mathbf{s}[k]$ is the \emph{suffix} from time~$k$ (i.e., $\mathbf{s}[k] = \{s_{t'}~|~t' \geq k\}$), $\mathbf{s}(i)$ is $i$-th term of a sequence~$\mathbf{s}$, where $\mathbf{s}(0)$ is the first term, $\mathbf{s}(last)$ is the last term and $|\mathbf{s}|$ is the cardinality of the sequence. For instance, we have $\mathbf{s}^{H}_{t}[t+2] =  s_{t+2} s_{t+3} \cdots$ and $\mathbf{s}^{H}_{t}(0) = s_t$.

The syntax of PrSTL is defined as
\begin{equation} \begin{split}
	\phi &::= \top \mid \neg \phi \mid \phi \wedge \phi \mid \phi_1~\calU_{[t_1,t_2]} \phi_2 \mid \calP_{\sim \lambda}[\psi]	\\
	\psi &::= \mu \mid \neg \psi \mid \psi \wedge \varphi \mid \calF_{[t_1,t_2]} \psi \mid \calG_{[t_1,t_2]} \psi
	,
\end{split} \end{equation}
where~$\top$ is a Boolean constant for `true', $\neg$ is a negation (`not'), $\wedge$ is a conjunction (`and') $\sim \in \{<,\leq,\geq,>\}$, $\lambda \in [0,1]$, $\calU$ is the `Until' temporal operator, $\calF$ is the `in Future' temporal operator, $\calG$ is the `Globally' temporal operator, $t_1,t_2 \in [0, \infty)$ such that $t_2 \geq t_1$, $\mu$ is a predicate over a real valued function of~$\mathbf{x}[t]$ (s.t. $\mu := r(\mathbf{x}[t])$ with~$r: \mathbb{R}^{n} \rightarrow \mathbb{B}$) and $\varphi \in \{ \phi, \psi \}$.
In this paper, we define two types of temporal logic formulas: \emph{event} and \emph{instance formulas}. An event formula~$\psi$ is specified over target beliefs given a run. Thus, a satisfaction of an event formula can be specified probabilistically. The probabilistic degree of satisfying an event formula~$\psi$ given a run~$\mathbf{x}$ and beliefs at time~$t$ is computed using a function~$\rmProb(\mathbf{x}, \psi, t)$. On the other hand, an instance formula~$\phi$ is defined over a sequence of truth values. Hence, a satisfaction can be known deterministically for a given sequence of agent states. 
We use $\calP_{\sim \lambda}[\cdot]$ operator to determine if the probability of satisfying a given event formula holds true for~$\sim \lambda$. For a synthesis problem where the objective is to either maximise or minimise the probability, we use special notations~$\calP_{max}$ and~$\calP_{min}$ respectively.

The semantics of PrSTL instance formulas are recursively defined as
\begin{equation} \begin{split}
	\mathbf{x}[t] \models \top, &\forall t	\\
	\mathbf{x}[t] \models \neg \phi \iff& \mathbf{x}[t] \not\models \phi \\
	\mathbf{x}[t] \models \phi_1 \wedge \phi_2 \iff& \mathbf{x}[t] \models \phi_1~\text{and}~\mathbf{x}[t] \models \phi_2	\\
	\mathbf{x}[t] \models \phi_1~\calU_{[t_1,t_2]} \phi_2 \iff& \exists t' \in [t_1,t_2]~\text{s.t.}~ \mathbf{x}_{t'} \models \phi_2~\text{and}	\\ &\forall t'' \in [t_1, t'-1], \mathbf{x}_{t''} \models \phi_1 \\
	\mathbf{x}[t] \models \calP_{\sim \lambda}[\psi] \iff& \rmProb(\mathbf{x}, \psi, t) \sim \lambda
	,
\end{split} \end{equation}

The satisfaction of a given event formula is measured probabilitically as
\begin{equation} \begin{split} \label{eqn:prob}
	\rmProb(\mathbf{x}, \mu, t) &= f^{\mu}(x_t)\\
	\rmProb(\mathbf{x}, \neg \psi, t) &= 1 - \rmProb(\mathbf{x}, \psi, t)	\\
	\rmProb(\mathbf{x}, \psi_1 \wedge \psi_2, t) &= \rmProb(\mathbf{x}, \psi_1, t) \cdot \rmProb(\mathbf{x}, \psi_2, t) \\
	\rmProb(\mathbf{x}, \psi \wedge \phi, t) &= \begin{cases}
			\rmProb(\mathbf{x}, \psi, t) & \text{if $\mathbf{x}[t] \models \phi$},	\\
			0 & \text{otherwise,}
		\end{cases}	\\
	\rmProb(\mathbf{x}, \calG_{[t_1,t_2]} \psi, t) &= \prod_{t' \in [t_1,t_2]} \rmProb(\mathbf{x}, \psi, t')	\\
	\rmProb(\mathbf{x}, \calF_{[t_1,t_2]} \psi, t) &= 1 - \prod_{t' \in [t_1,t_2]} (1 - \rmProb(\mathbf{x}, \psi, t'))
	,
\end{split} \end{equation}
where $f^{\mu}: \mathbb{R}^n \times \mathbb{R}^{n_{\mu}} \rightarrow \mathbb{R}$. Given a run~$\mathbf{x} = \{ x_0, \cdots, x_t, \bar{x}_{t+1}, \cdots \}$, $\rmProb(\mathbf{x}, \mu, k) = z^{\mu}_k \in \{ 0, 1 \}$ if $k \leq t$. Otherwise, $\rmProb(\mathbf{x}, \mu, k) = \mathbb{P}(\hat{x}^{\mu}_k \mid \mathbf{Z}^{\mu}_k)$.

From the existing operators, additional operators can be derived:
\begin{equation} \begin{split}
	\varphi_1 \vee \varphi_2 &= \neg (\neg \varphi_1 \wedge \neg \varphi_2)	\\
	\varphi_1 \Rightarrow \varphi_2 &= \neg \varphi_1 \vee \varphi_2	\\
	\calF_{[t_1,t_2]} \varphi &= \top~\calU_{[t_1,t_2]} \varphi	\\
	\calG_{[t_1,t_2]} \varphi &= \neg \calF_{[t_1,t_2]} \neg \varphi
	,
\end{split} \end{equation}
where $\Rightarrow$ is an implication (i.e., \emph{if $\varphi_1$, then $\varphi_2$}), $\calF$ is a temporal operator for 'sometime in future' (eventually), and $\calG$ is a temporal operator for `globally' (always).

Every PrSTL formula has a \emph{horizon length} denoted as~$hrz(\varphi) \in \mathbb{N}^0$ (i.e., non-negative integer)~\cite{dokhanchi2014on}. The horizon length is the minimum length in time of a signal (i.e., a run of an agent) required to evaluate the signal against a given specification~$\varphi$. The horizon length can be computed resursively as
\begin{equation} \begin{split}
	hrz(\mu) &= 0	\\
	hrz(\neg \varphi) &= hrz(\varphi)	\\
	hrz(\varphi_1 \wedge \varphi_2) &= \max \{ hrz(\varphi_1), hrz(\varphi_2) \}	\\
	hrz(\varphi_1\calU_{[t_1,t_2]} \varphi_2) &= t_2 + \max \{ hrz(\varphi_1) - 1, hrz(\varphi_2) \}
	.
\end{split} \end{equation}

Using PrSTL, Examples~\ref{exa:surveillance} and~\ref{exa:search} can be re-written as the following.
\begin{example} [continues=exa:surveillance]
	The surveillance mission can be re-written as
	\begin{equation} \begin{split} \label{eqn:prstl_surv}
			\calG_{[0, 30]} ( \calF_{[0, 40]} \mu_1 \wedge \calF_{[0, 40]} \mu_2 \wedge \calF_{[0, 40]} \mu_3 )
		,
	\end{split} \end{equation}
	where~$\mu_1$, $\mu_2$ and~$\mu_3$ are the predicates for targets in the area. 
	Over $30$ minutes, each target has to be located repeatedly every $40$ minutes.
	The horizon length of the mission is~$70$.
\end{example}
\begin{example} [continues=exa:search]
	The search mission can be re-written as
	\begin{equation} \begin{split} \label{eqn:prstl_search}
			\calF_{[0,60]} \mu_{Tom} \wedge \calG_{[0,60]} (\calP_{=1}[\mu_{Tom}] \Rightarrow \calF_{[0,30]} \mu_{Jerry}) 
		,
	\end{split} \end{equation}
	where~$\mu_{Tom}$ and~$\mu_{Jerry}$ are the predicates for Tom and Jerry respectively. In this mission, Tom has to be found in $60$ minutes. Whenever Tom is located, Jerry needs to be found in $40$ minutes. The horizon length of the mission is~$90$.
\end{example}



\section{RECEDING HORIZON SYNTHESIS WITH FORWARD SEARCH} \label{sec:synthesis}

\begin{table}[t!]
	\caption{Evaluations of a formula~$\calG_{[0,1]} \calF_{[0, 3]} \mu$ over trajectories~$\mathbf{x}$, $\mathbf{x}^3$ and~$\mathbf{x}^5$. Approximated numbers are shown in shaded cells. Note that $\psi_{\calF} = \calF_{[0,3]} \mu$ and $\psi_{\calG} = \calG_{[0,1]} \psi_{\calF}$.}
	\centering
	\begin{tabular}{l|cccccc}
		& \multicolumn{6}{c}{Time~$t$}	\\
		&	$0$	&	$1$	&	$2$	&	$3$	&	$4$	&	$5$	\\
		\hline \hline
		$\rmProb(\mathbf{x}, \mu, t)$	&	0.8	&	0.7	&	0.5	&	0.6	&	0.6	&	0.7	\\
		$\rmProb(\mathbf{x}, \psi_{\calF}, t)$	& 0.988 & 0.976 & 0.976 & $\cdots$ & $\cdots$ & $\cdots$	\\
		$\rmProb(\mathbf{x}, \psi_{\calG}, t)$	& 0.964 & 0.953 & $\cdots$ & $\cdots$ & $\cdots$ & $\cdots$	\\
		
		\hline
		$\rmProb'(\mathbf{x}^3, \mu, t)$	&	0.8	&	0.7	&	0.5	&	0.6	& N/A & N/A 	\\
		$\rmProb'(\mathbf{x}^3, \psi_{\calF}, t)$	& 0.988 &\graycell 0.94 &\graycell 0.8 &\graycell 0.6 & N/A & N/A	\\
		$\rmProb'(\mathbf{x}^3, \psi_{\calG}, t)$	&\graycell 0.929 &\graycell 0.752 &\graycell 0.48 &\graycell 0.6 & N/A & N/A 	\\
		
		\hline
		$\rmProb'(\mathbf{x}^5, \mu, t)$	&	0.8	&	0.7	&	0.5	&	0.6	&	0.6	& 0.7	\\
		$\rmProb'(\mathbf{x}^5, \psi_{\calF}, t)$	& 0.988 & 0.976 & 0.976 &\graycell 0.952 &\graycell 0.88 &\graycell 0.7	\\
		$\rmProb'(\mathbf{x}^5, \psi_{\calG}, t)$	& 0.964 & 0.953 &\graycell 0.929 &\graycell 0.838 &\graycell 0.616 &\graycell 0.7	\\
		
	\end{tabular}
	\label{tab:relax}
\end{table}

To solve Problem~\ref{problem:finite} in an efficient manner, we propose an algorithm using \emph{forward search} and RHC method. The algorithm assumes that time bounds on temporal operators start from zero (i.e., $\calF_{[0, \tau]} \varphi$ and $\calG_{[0, \tau]} \varphi$).

The algorithm works as follows. 
Given a sequence of past agent states~$\mathbf{X}_t$ at time~$t$, we iteratively apply all the control inputs to generate a set of \emph{candidate trajectories}~$\mathbf{C}^t_k$ at $k$-th iteration. A candidate trajectory is a run that consists of past agent states and future agent states. 
Starting from~$\mathbf{C}^t_0 = \{ \{ \mathbf{X}_t \} \}$, we iteratively update the set of candidate trajectories as follows:
\begin{equation} \begin{split} \label{eqn:new_branch}
	\mathbf{C}^t_{i+1} = \{ \{ \mathbf{c}, f( \mathbf{c}(last), u ) \} \mid \forall u \in \mathbf{U} \text{ and } \forall \mathbf{c} \in \mathbf{C}^t_{i} \}
	.
\end{split} \end{equation}
We repeat the process to generate a new set of candidate trajectories until we reach the end of horizon length~(i.e., $t+i = H$). However, the growth in the number of trajectories is not scalable in practice. Therefore we introduce a heuristically chosen constant~$N$ which is the maximum number of candidate trajectories. Thus, after each iteration, we compute the probabilistic degree of satisfaction using~$\rmProb$ and only maintain $N$-best trajectories for the next iteration. 
For a set of candidate trajectories~$\mathbf{C}^t_i$, we find a new set~$\tilde{\mathbf{C}}^t_i$ such that
\begin{equation} \begin{split} \label{eqn:prune}
	\tilde{\mathbf{C}}^t_i = \{ \mathbf{c} \in \mathbf{C}^t_i \mid \rmProb(\mathbf{c}, \psi, 0) \leq \rmProb(\mathbf{c}', \psi, 0), \forall \mathbf{c}' \in \mathbf{C}' \}
	,
\end{split} \end{equation}
where~$|\mathbf{C}'| = N$. We replace~$\mathbf{C}^t_i$ in~(\ref{eqn:new_branch}) with~$\tilde{\mathbf{C}}^t_i$ to calculate $\mathbf{C}^t_{i+1}$.
After we compute the next set of trajectories, we check if there exists only one branch from the initial state. If so, we stop and execute the corresponding control input leading to the branch. Formally speaking, we stop when the condition below is true for a set of candidate trajectories~$\tilde{\mathbf{C}}^t_i = \{ \mathbf{c}_1, \mathbf{c}_2, \cdots, \mathbf{c}_N \}$:
\begin{equation} \begin{split}
	\mathbf{c}_i(t+1) = \mathbf{c}_j(t+1), \forall i, j \leq N
	.
\end{split} \end{equation}
This is because we use RHC method in which only the first control input from the sequence is important. Therefore, computing any further is not computationally beneficial. 

\begin{figure}[t]
	\centering
	\subfloat[1st iteration]{\includegraphics[scale=0.75]{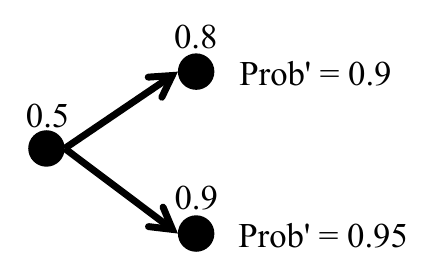}}
	\subfloat[2nd iteration]{\includegraphics[scale=0.75]{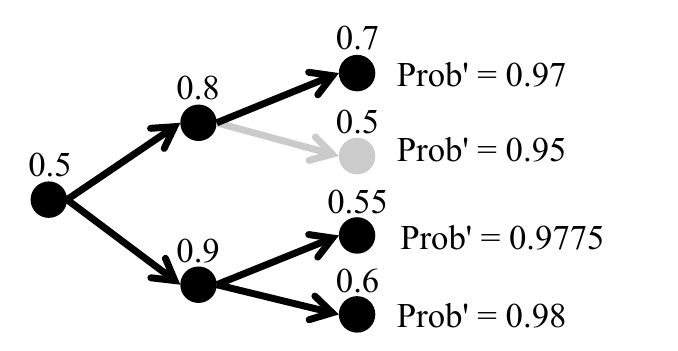} \label{fig:branch-2}}
	
	\subfloat[3rd iteration]{\includegraphics[scale=0.75]{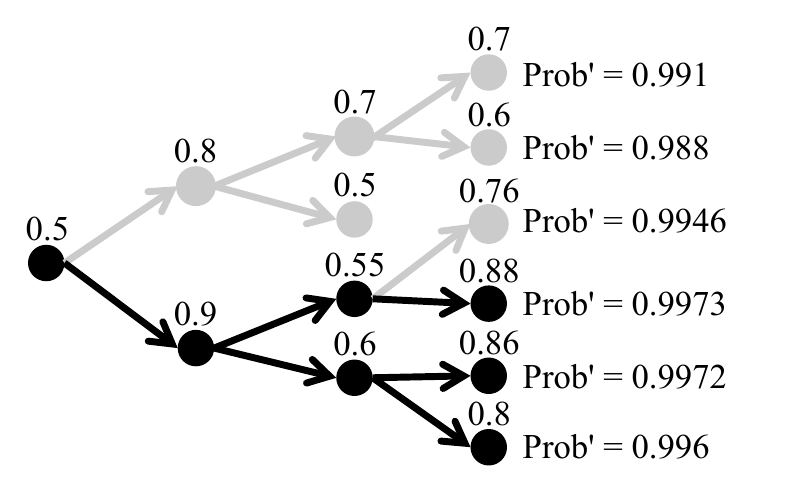} \label{fig:branch-3}}
	\caption{Demonstration example for the forward search algorithm with a PrSTL formula is~$\calF_{[0, 5]} \mu$, $N = 3$ and $|\mathbf{U}| = 2$. Each node represents a state of a agent where the number on every node is the probability that $\mu$ is satisfied in the state. The overall relaxed satisfaction probabilities are shown for resulting candidate trajectories after each iteration.}
	\label{fig:branch}
\end{figure}

In order to evaluate a given trajectory over a PrSTL formula, the length of the trajectory has to be equal or greater than the horizon length of the formula as described in Sec.~\ref{sec:prstl}. However, the proposed synthesis algorithm requires an evaluation when the length is shorter than the horizon length. Hence, we propose a relaxation to evaluate short trajectories. 
Given a sequence of states with finite length~$\mathbf{x}^n$, the approximated probability of satisfaction for a temporal operator is re-written as
\begin{equation} \begin{split}
	\rmProb'(\mathbf{x}^n, \mathcal{T}_{[0, \tau]} \varphi, t) = \rmProb(\mathbf{x}', \mathcal{T}_{[0, \min(\tau, n)]} \varphi, t)
\end{split} \end{equation}
where~$\mathcal{T} \in \{ \calG, \calF \}$ and $\mathbf{x}^n$ is a prefix of the infinite sequence~$\mathbf{x}'$. We replace~$\rmProb$ in (\ref{eqn:prune}) with~$\rmProb'$.
Suppose we have a PrSTL formula~$\calG_{[0,1]} \calF_{[0,3]} \mu$, the satisfaction probability over a set of example trajectories~($\mathbf{x}$, $\mathbf{x}^3$ and~$\mathbf{x}^5$) are shown in Tab.~\ref{tab:relax} where the probabilities in shades are approximated.

In Fig.~\ref{fig:branch}, we illustrate an example of a PrSTL formula~$\calF_{[0, 5]} \mu$ over three iterations where~$N = 3$ and~$|\mathbf{U}| = 2$. After the first iteration, we have two candidate trajectories. At the next iteration, we again apply control inputs to every candidate trajectories and obtain four new trajectories as shown in Fig.~\ref{fig:branch-2}. As the number of trajectories is greater than the limit, we calculate the relaxed satisfaction probability for each trajectory and remove the least satisfying branch. We repeat the same in the third iteration. As there exists only one branch from the starting state, we terminate the process and execute the corresponding control.

\section{DISCUSSIONS} \label{sec:analysis}



Using forward search and RHC, we have gained a significant improvement in efficiency in solving the problem. This is achieved by limiting the number of candidate trajectories. If the number were not limited, the time complexity would be proportional to~$|\mathbf{U}|^{H-t}$ at time~$t$ which are not scalable in practice where agent operates over a long mission horizon. With the limiting constant~$N$, the complexity is reduced to~$|\psi| \cdot |N| \cdot |\mathbf{U}| \cdot (H-t)$ where~$|\psi|$ is the size of formula, $H$ is the horizon length.
This is because we apply $|\mathbf{U}|$-number of control inputs to $|N|$-number of candidate trajectories over $H-t$ iterations where each newly created trajectory is approximately evaluated $|\psi|$ times. The overall time complexity of the mission is~$|\psi| \cdot |N| \cdot |\mathbf{U}| \cdot H^2$. 


Since we limit the number of candidate trajectories, the completeness of the algorithm in finding the optimal solution with respect to satisfaction probability is not assured. However, our algorithm has gained a significant improvement in time complexity in return. In the following section, we show that the algorithm runs fast enough for an online synthesis in the presence of a changing belief space.


\begin{figure}[!t]
	\centering
	\subfloat[$t = 1$]{\includegraphics[width=0.48\columnwidth]{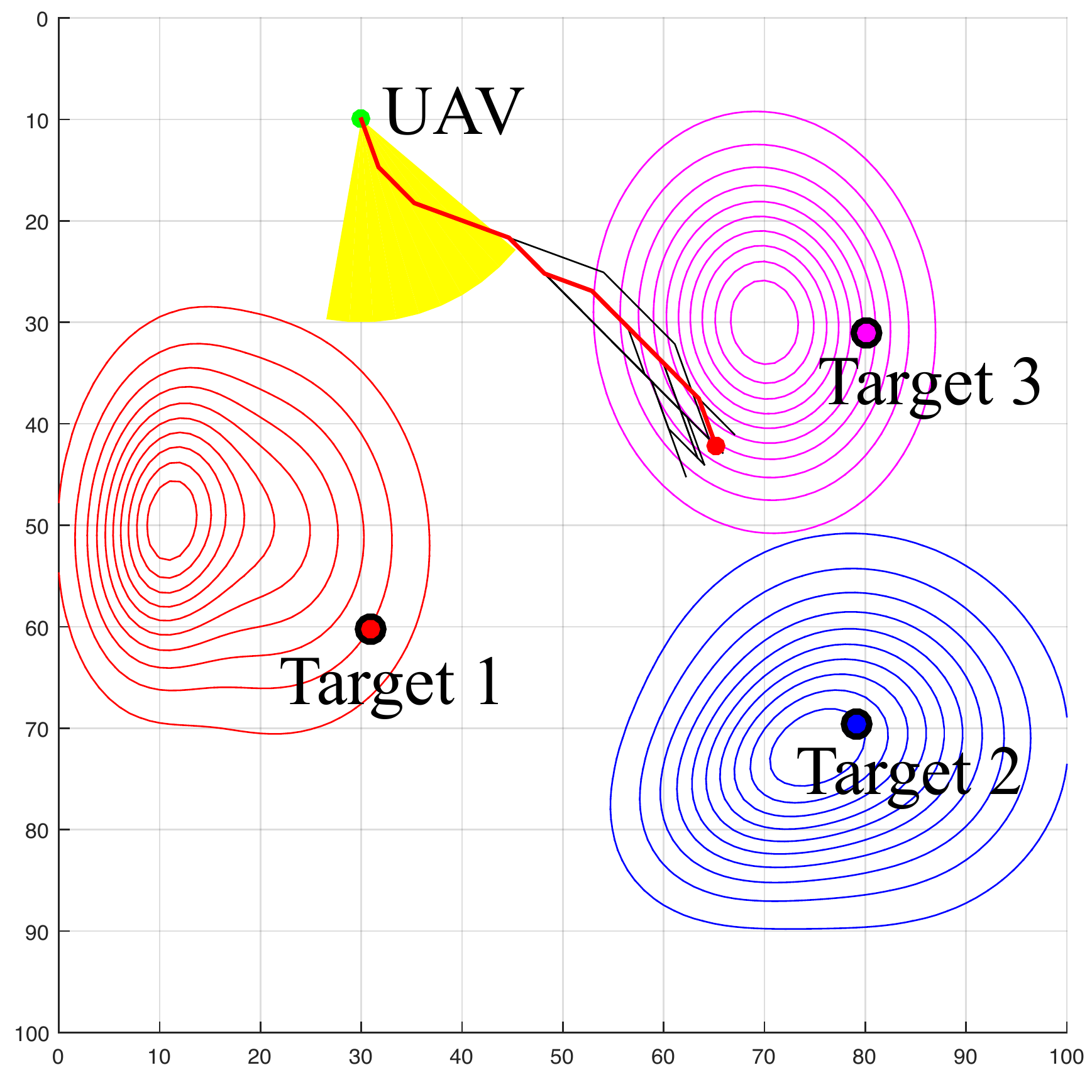} \label{fig:exec_surv_1}}
	\subfloat[$t = 12$]{\includegraphics[width=0.48\columnwidth]{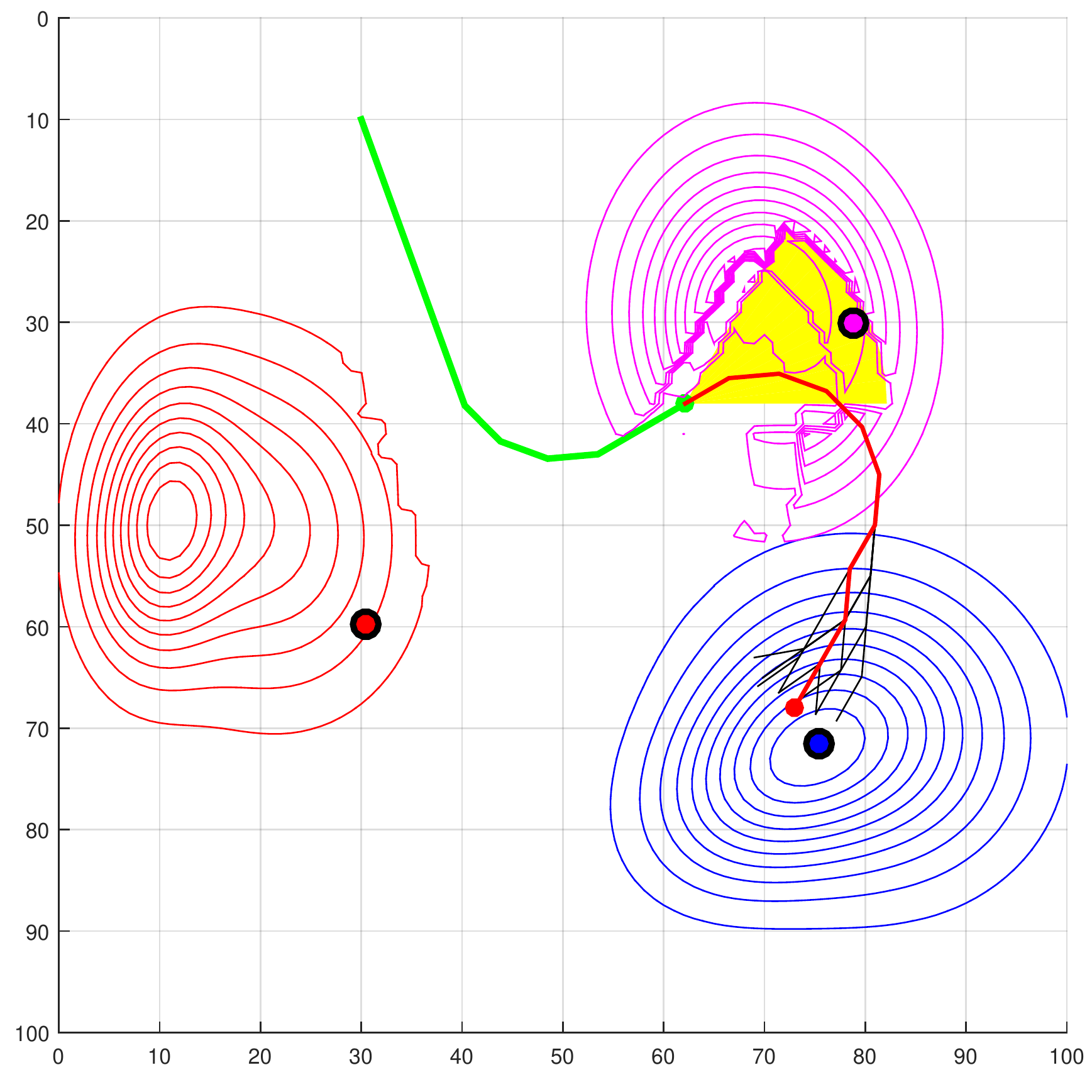}}
	
	\subfloat[$t = 13$]{\includegraphics[width=0.48\columnwidth]{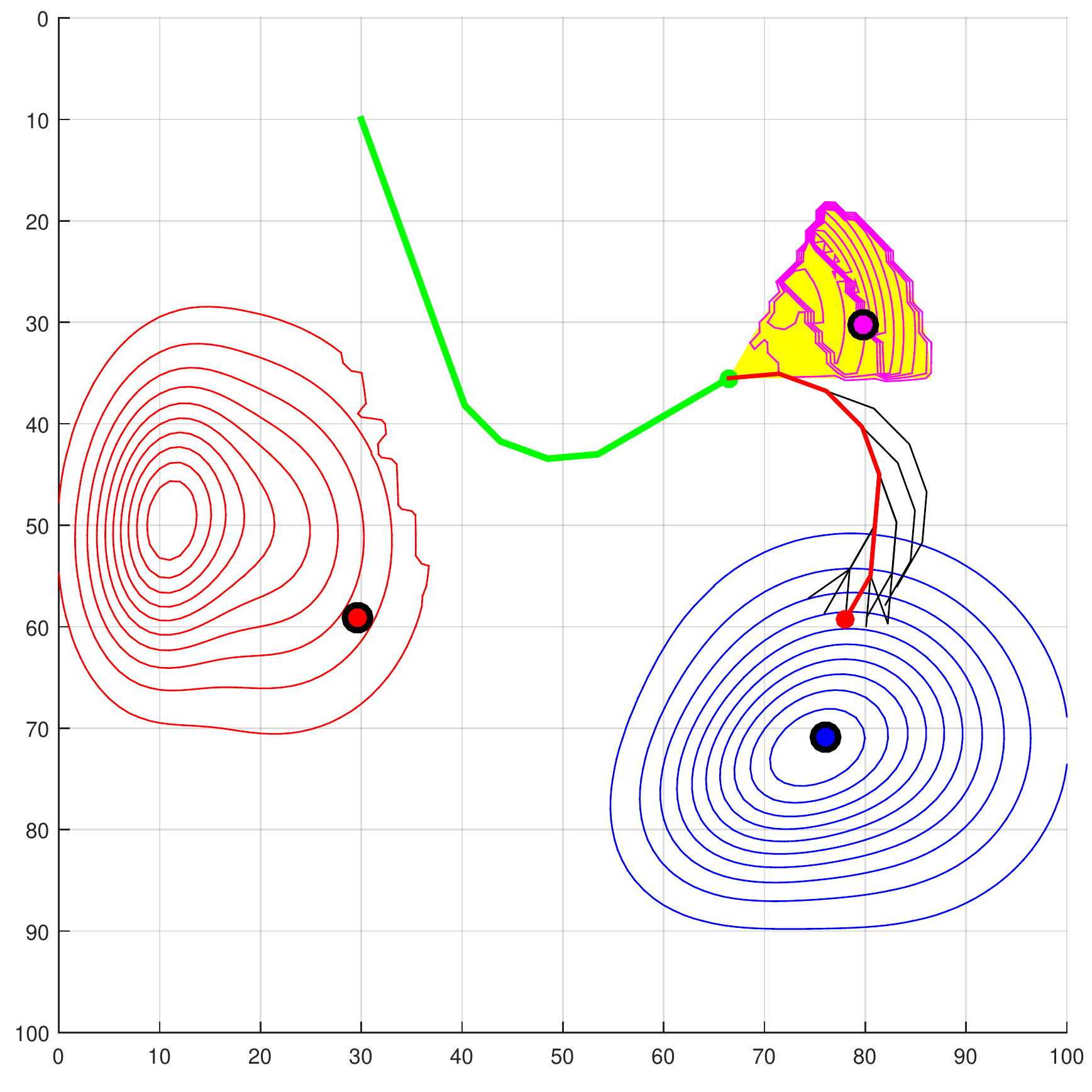}}
	\subfloat[$t = 18$]{\includegraphics[width=0.48\columnwidth]{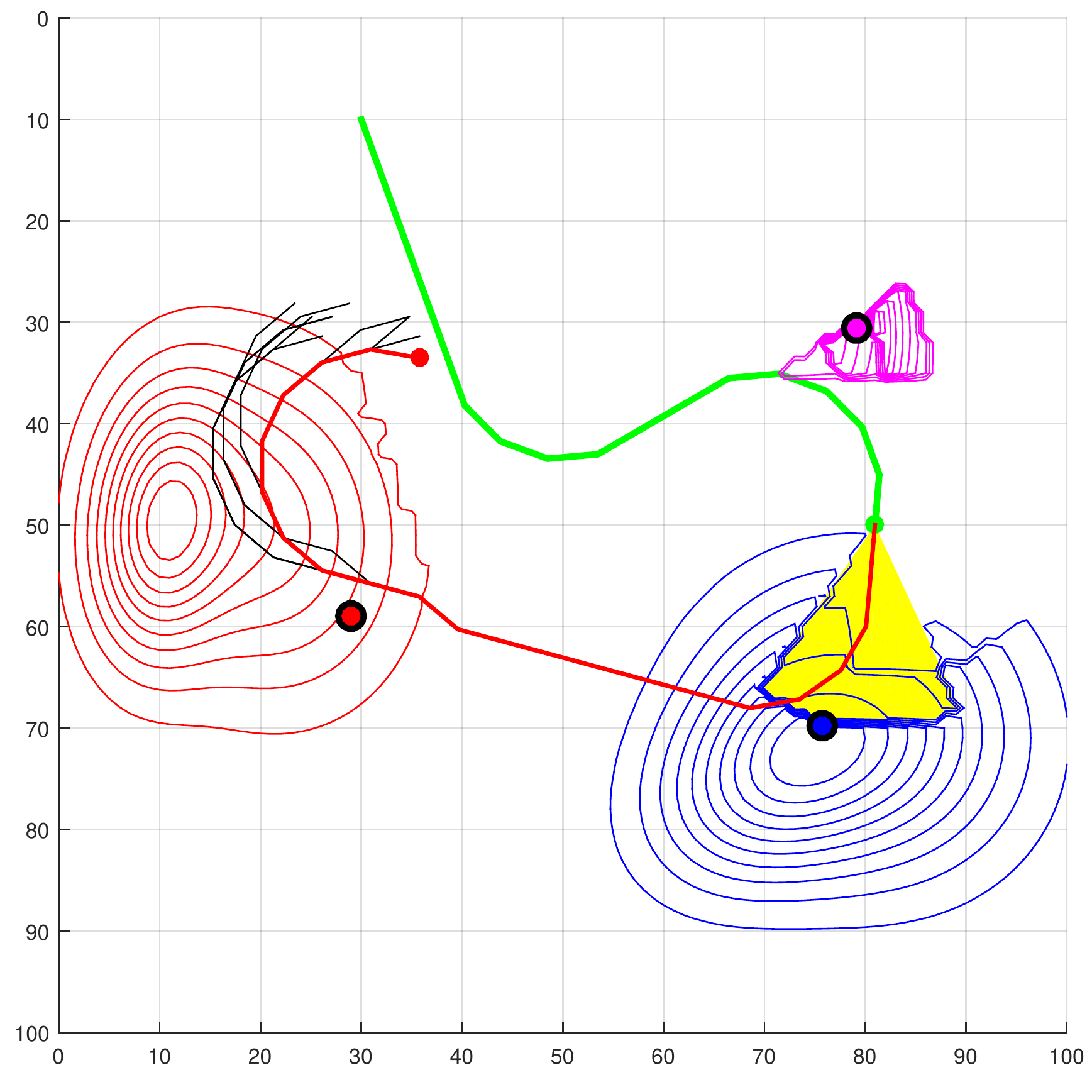}}
	
	\subfloat[$t = 22$]{\includegraphics[width=0.48\columnwidth]{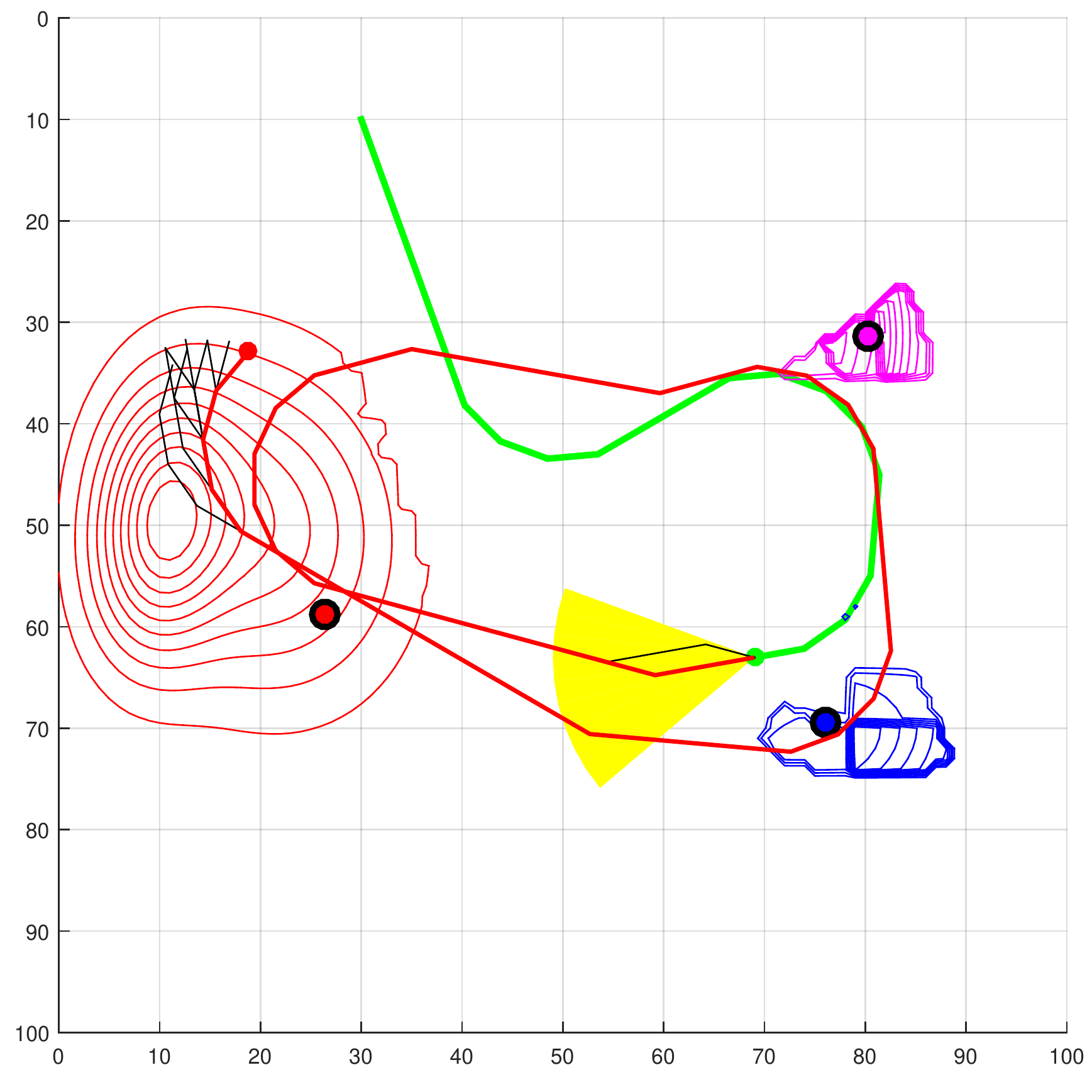}}
	\subfloat[$t = 27$]{\includegraphics[width=0.48\columnwidth]{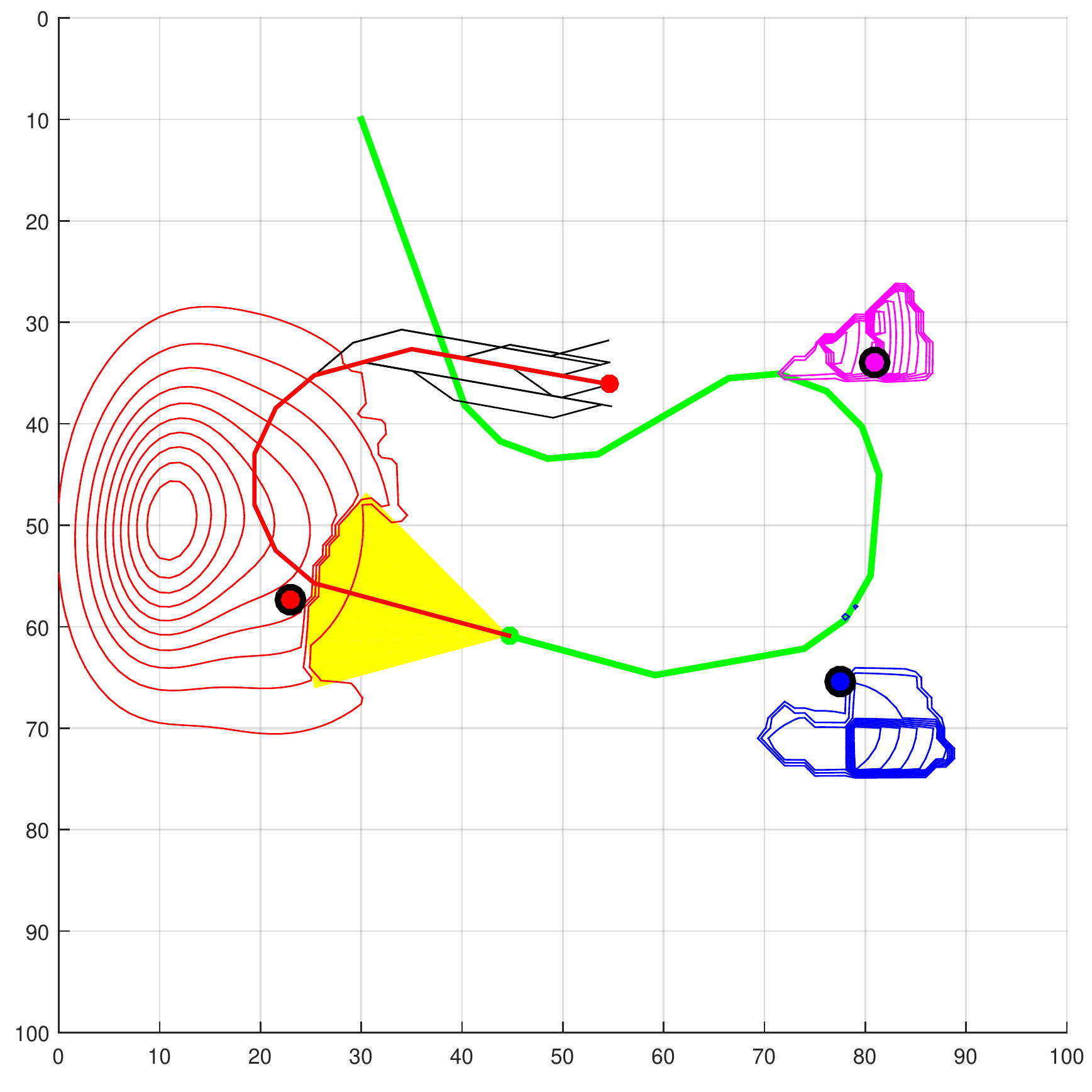}}
	
	\subfloat[$t = 28$]{\includegraphics[width=0.48\columnwidth]{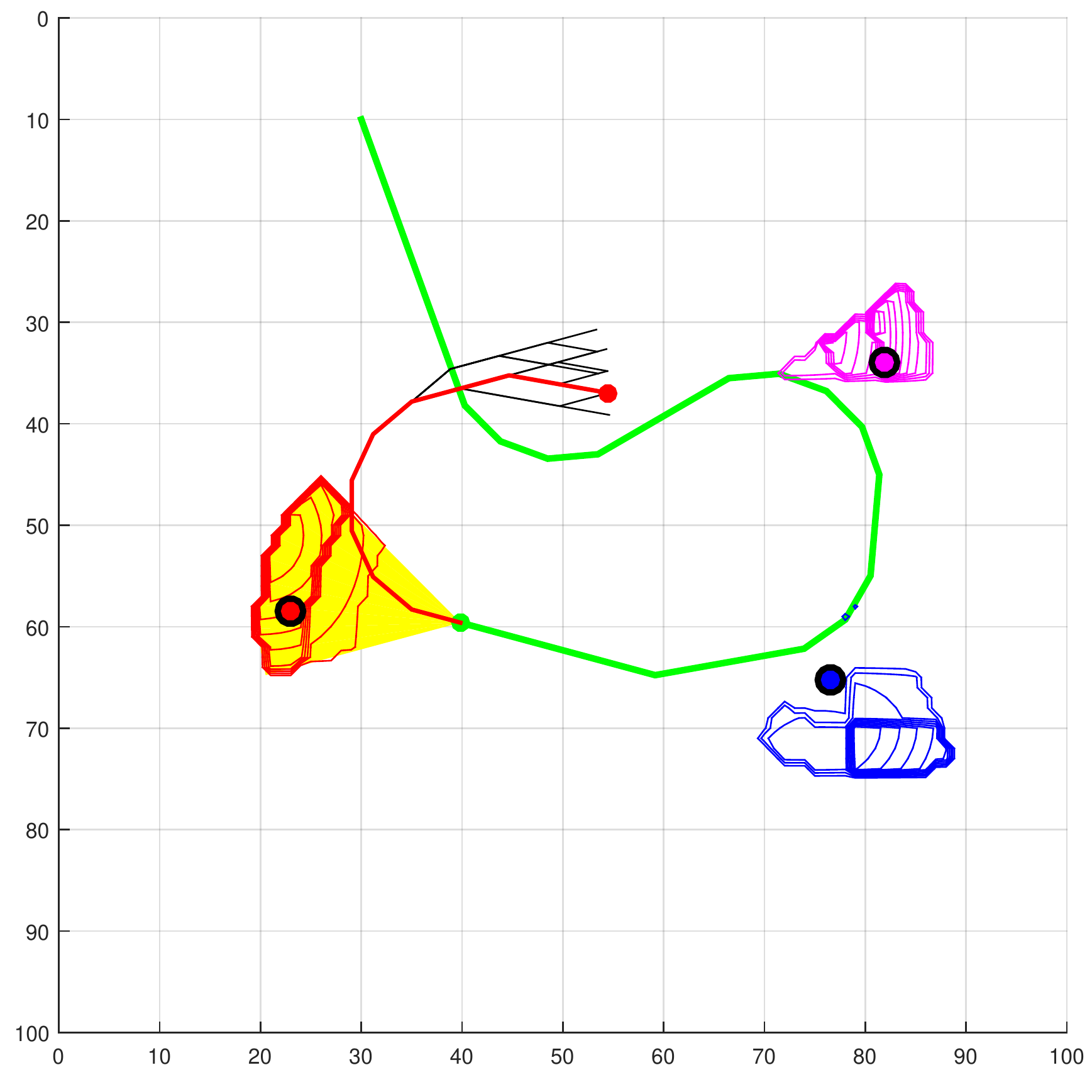}}
	\subfloat[$t = 59$ (end)]{\includegraphics[width=0.48\columnwidth]{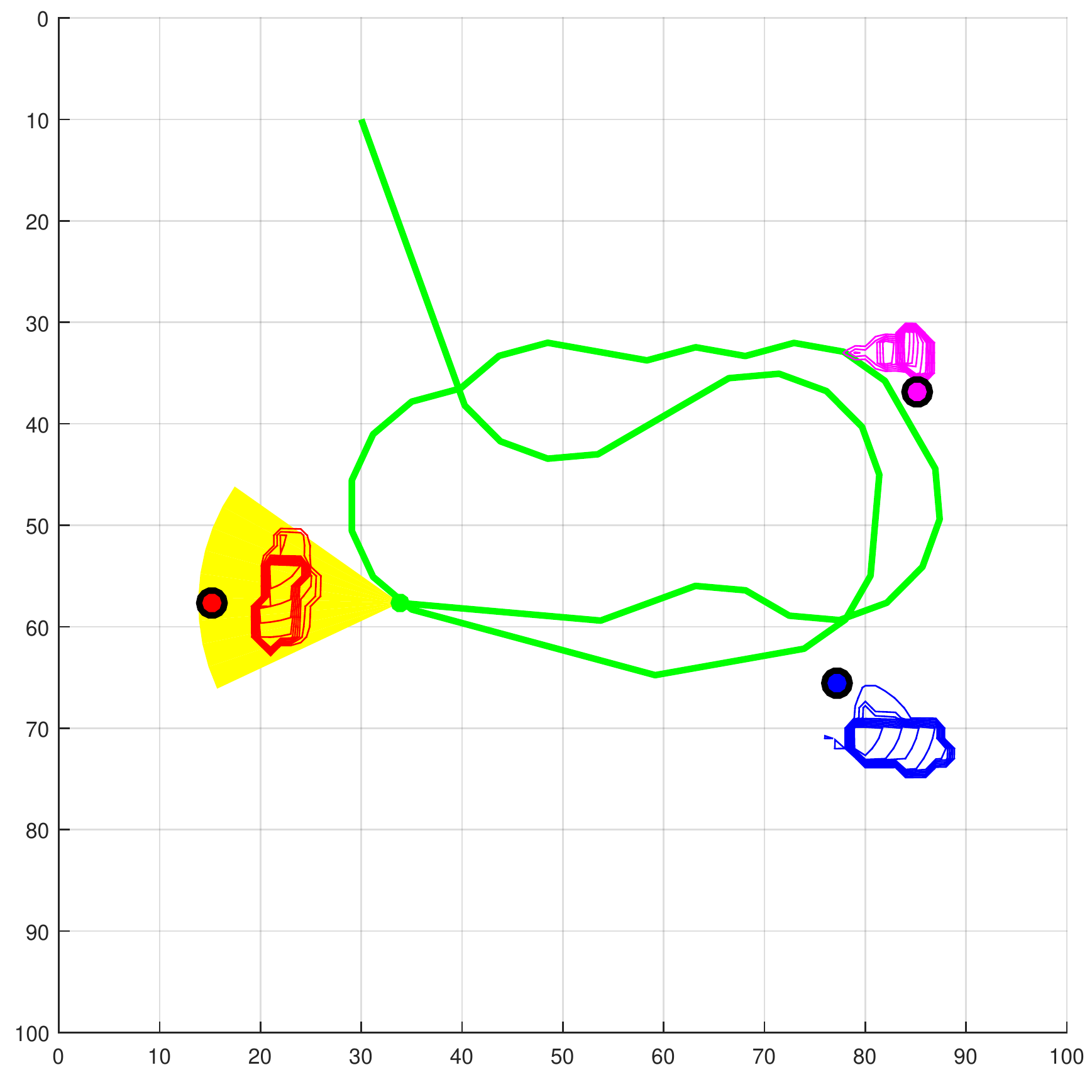}}
	
	\caption{Resulting trajectories for the surveillance mission (Example~\ref{exa:surveillance}). Targets are shown in contours of different colours. A synthesis of a control is based on the current beliefs over the targets. }
	\label{fig:exec_surv}
\end{figure}
\begin{figure}[!t]
	\centering
	\subfloat[$t = 1$] {\includegraphics[width=0.48\columnwidth]{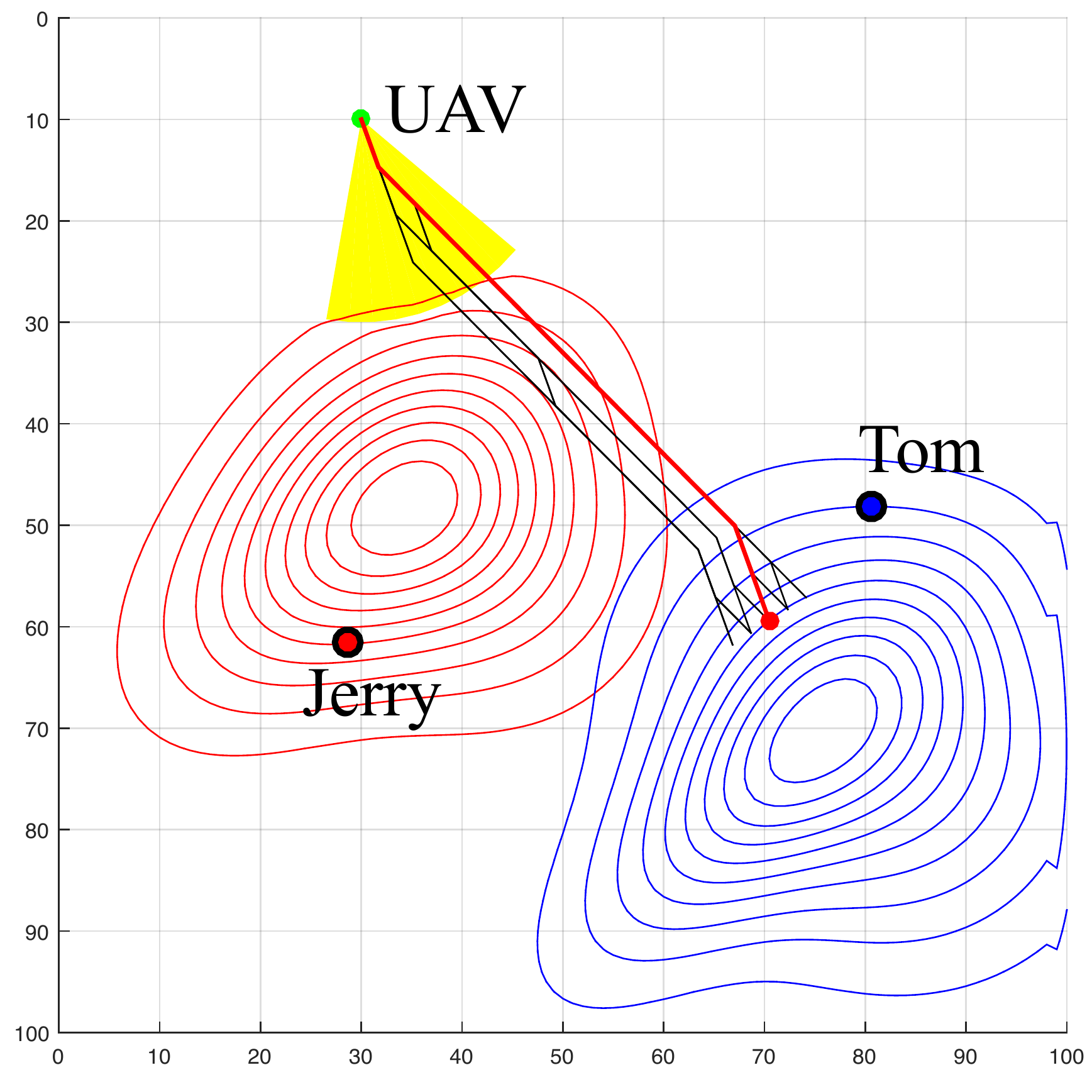} \label{fig:exec_a_then_b_1}}
	\subfloat[$t = 10$]{\includegraphics[width=0.48\columnwidth]{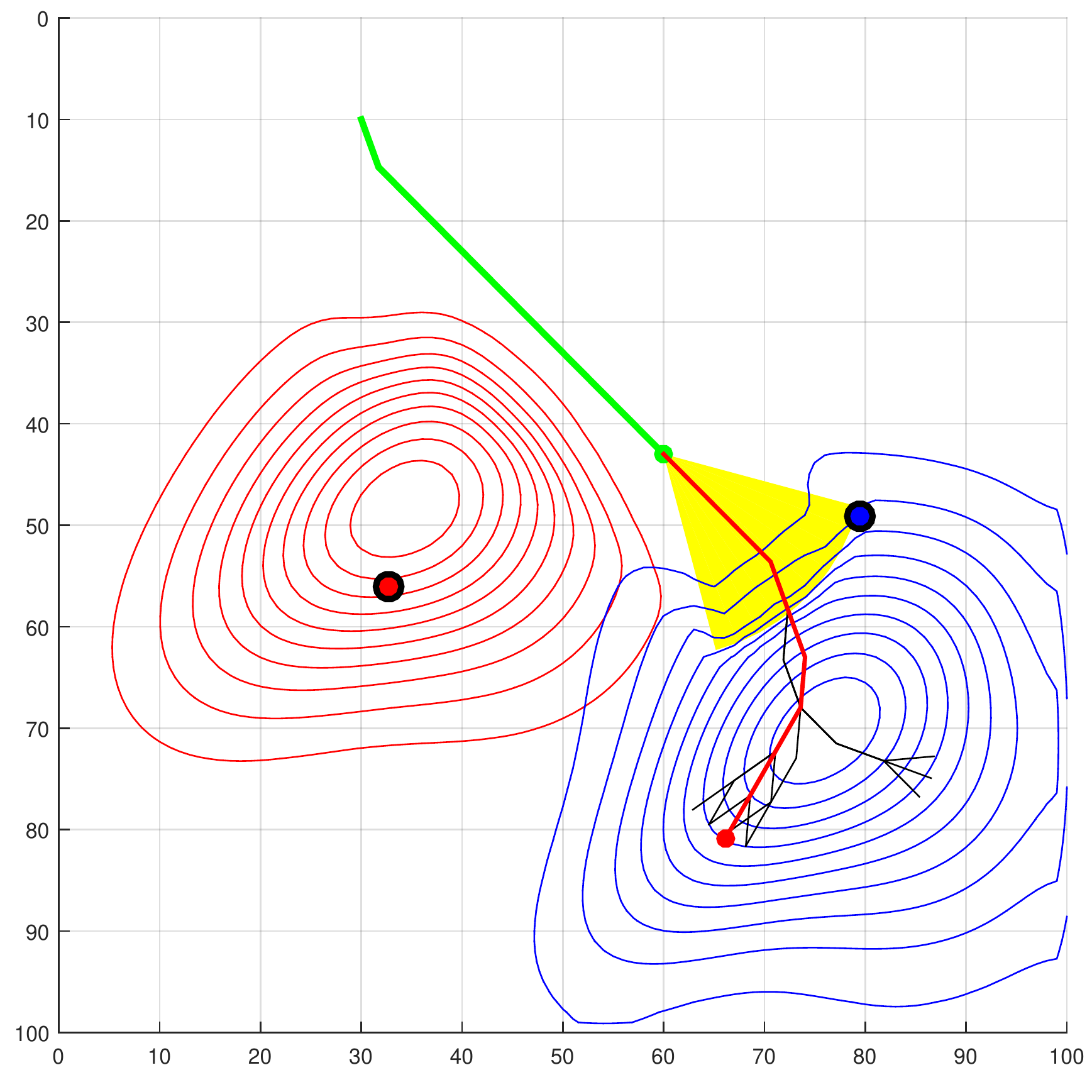}}
	
	\subfloat[$t = 13$]{\includegraphics[width=0.48\columnwidth]{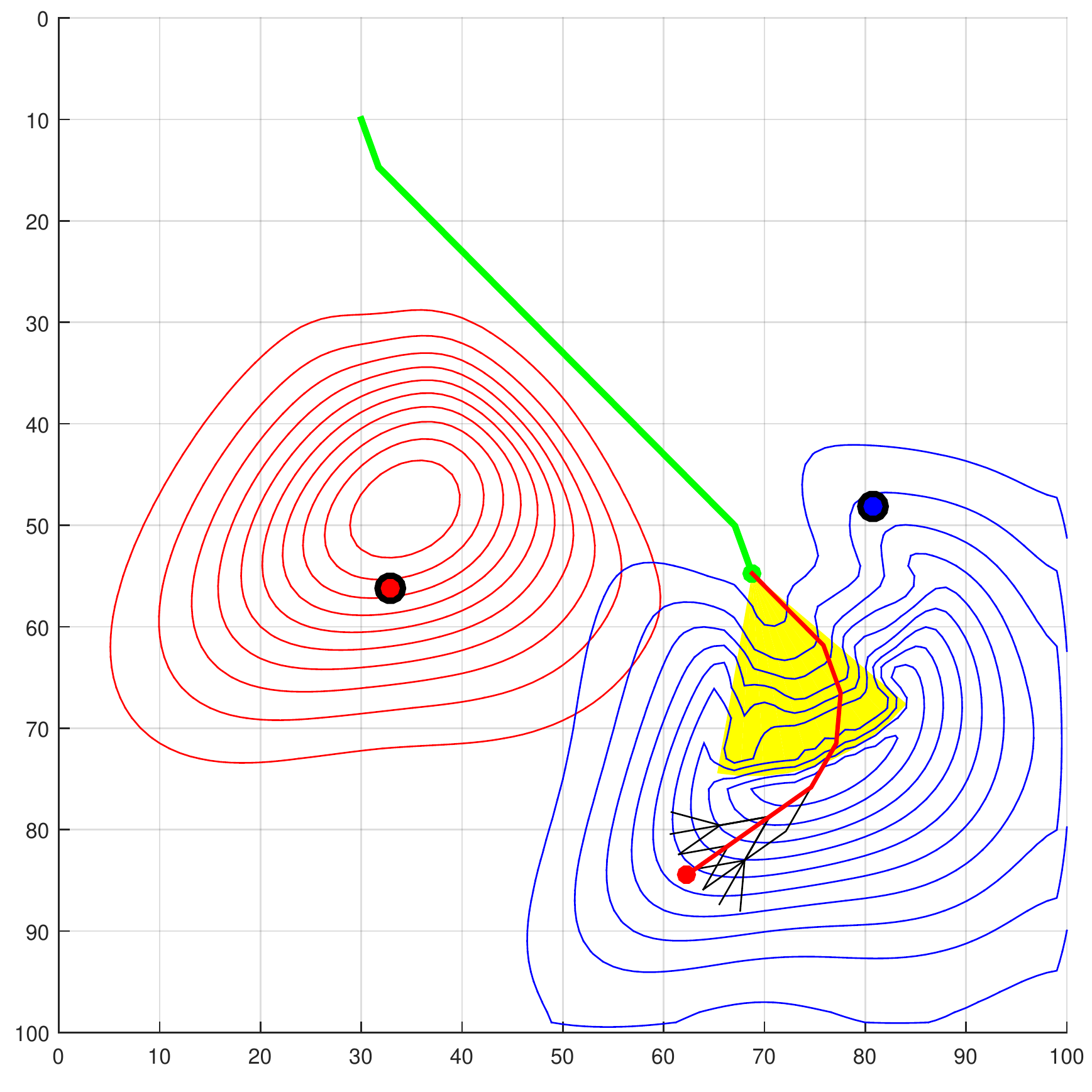}}
	\subfloat[$t = 19$]{\includegraphics[width=0.48\columnwidth]{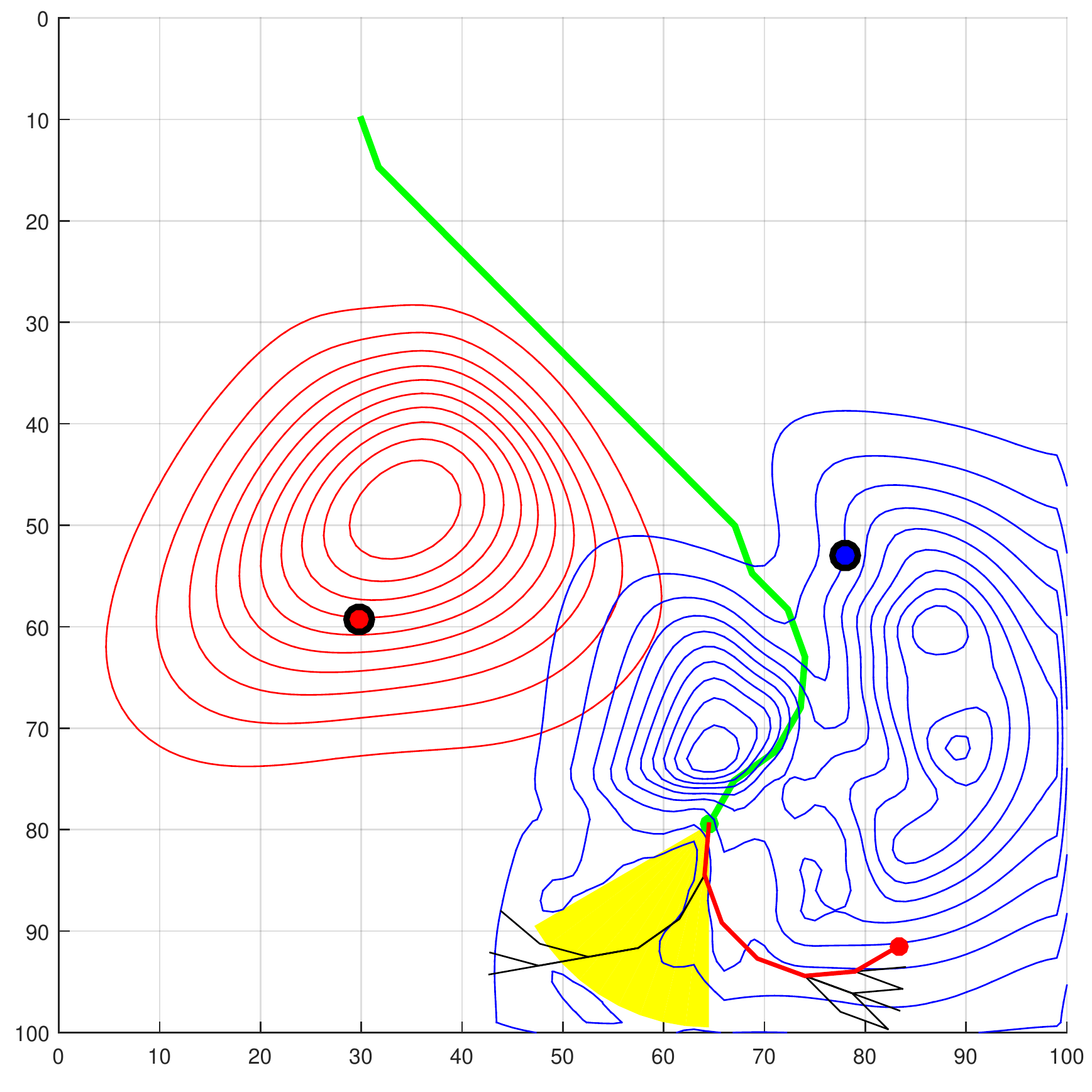} \label{fig:exec_a_then_b_19}}
	
	\subfloat[$t = 26$]{\includegraphics[width=0.48\columnwidth]{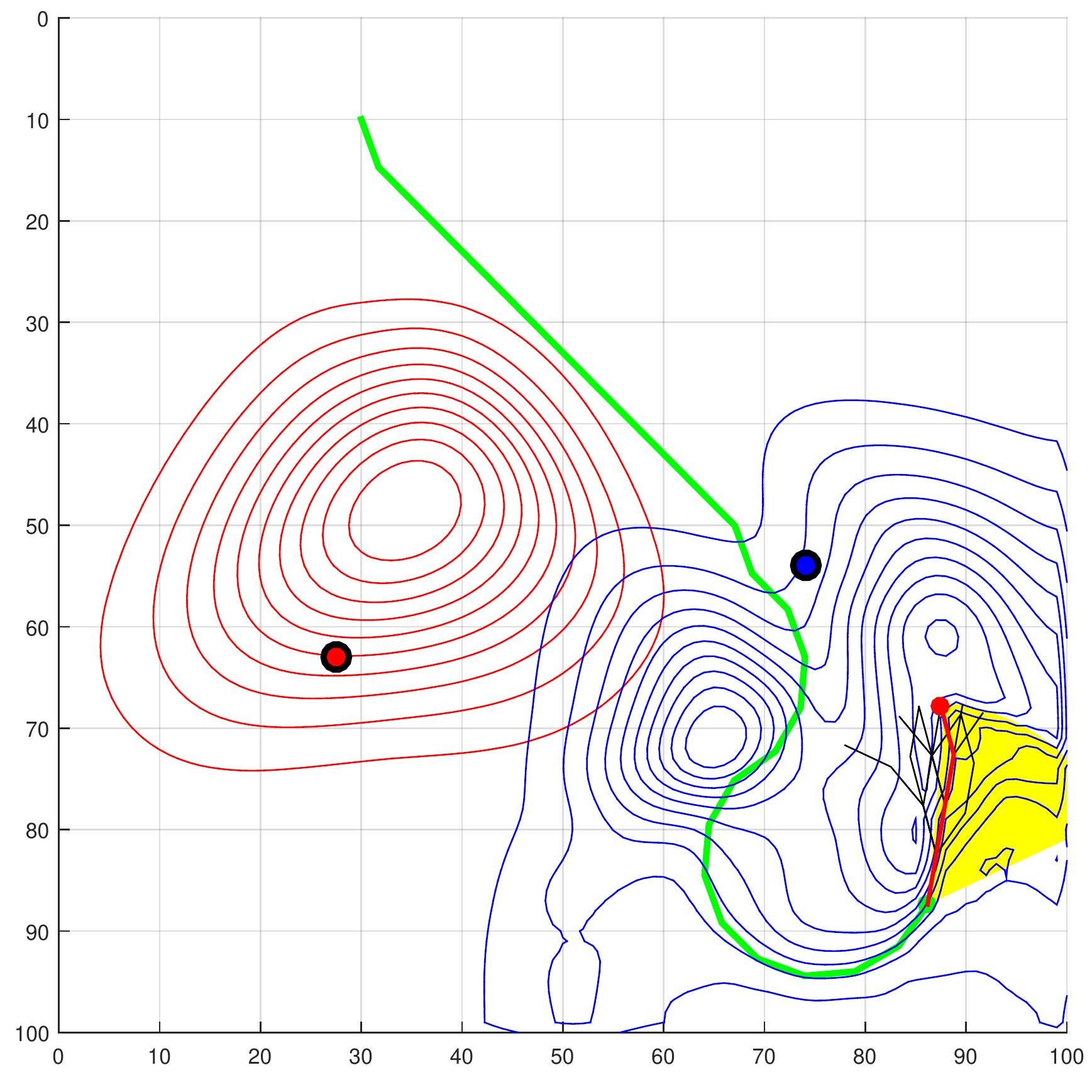} \label{fig:exec_a_then_b_26}}
	\subfloat[$t = 29$]{\includegraphics[width=0.48\columnwidth]{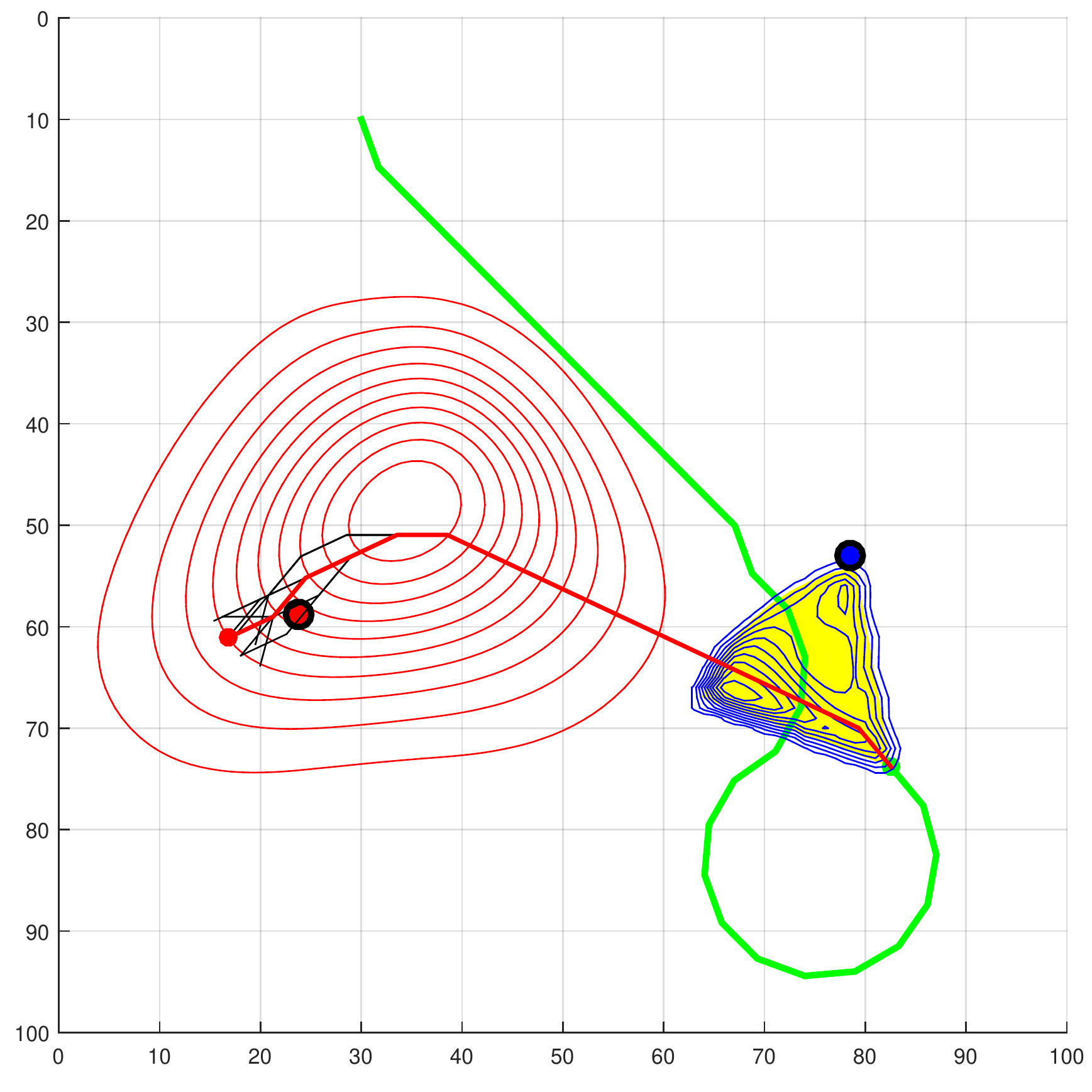}}
	
	\subfloat[$t = 35$]{\includegraphics[width=0.48\columnwidth]{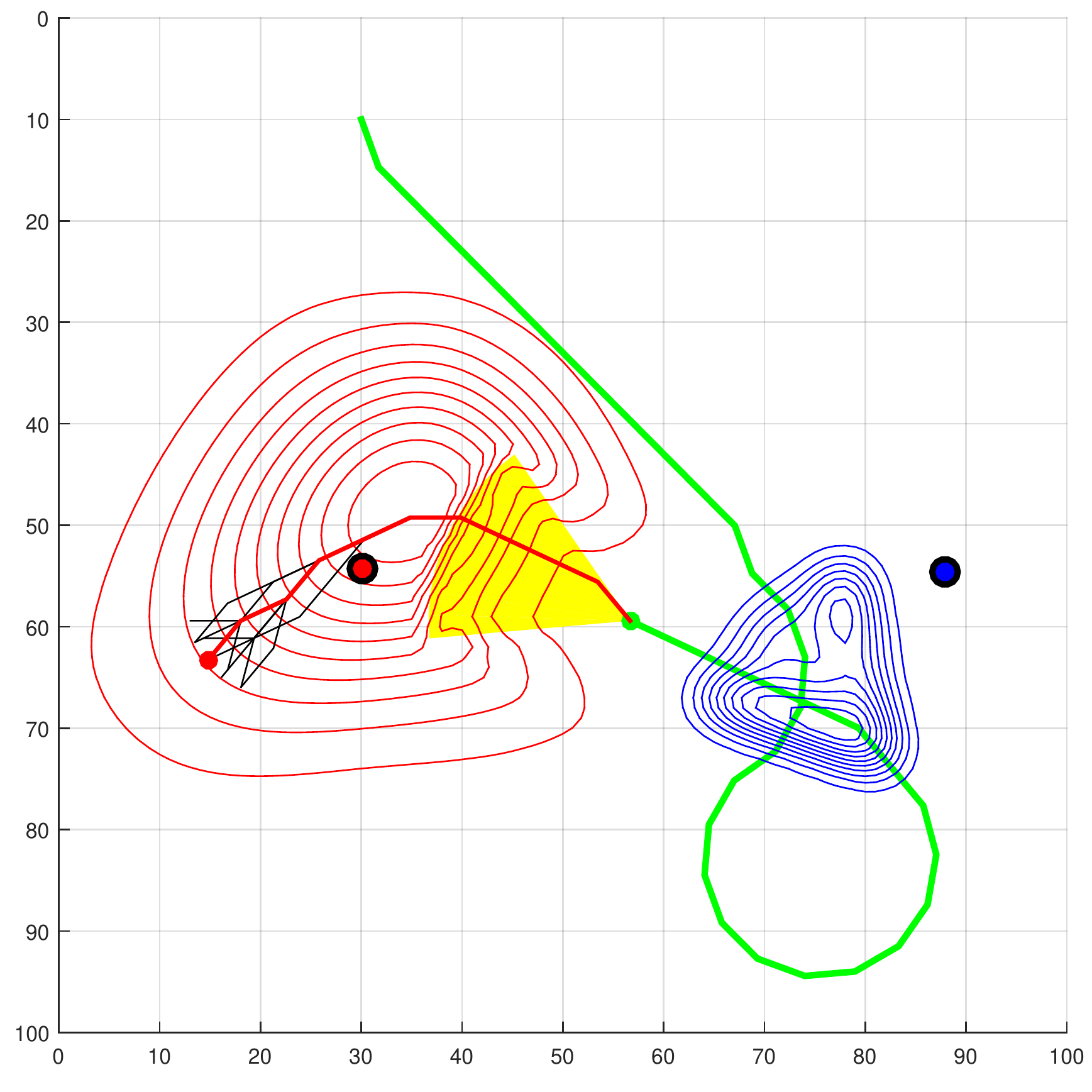}}
	\subfloat[$t = 37$ (end)] {\includegraphics[width=0.48\columnwidth]{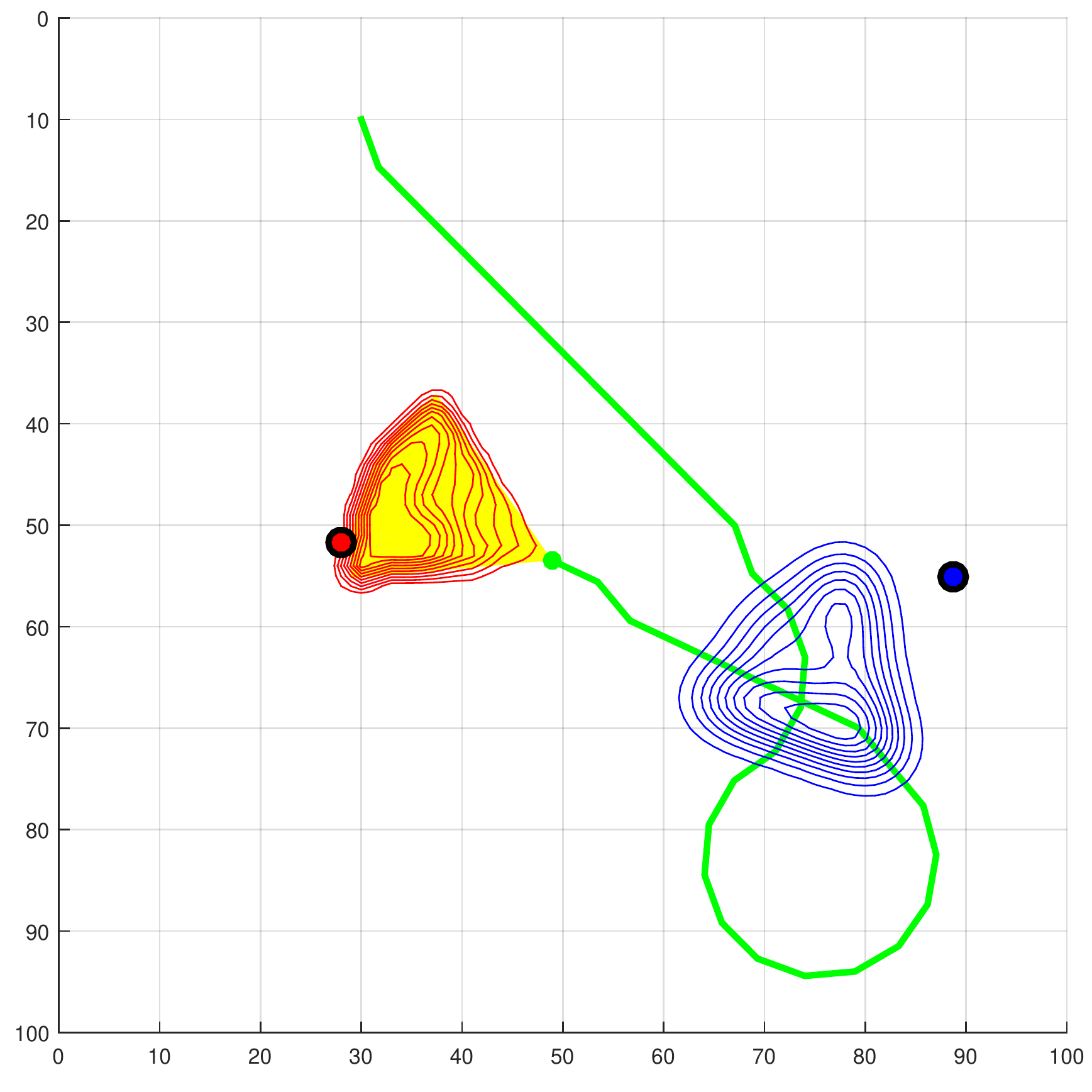}}
	
	\caption{Resulting trajectories for the prioritised search mission (Example~\ref{exa:search}). Blue and red contours represent the belief over Tom and Jerry, respectively.}
	\label{fig:exec_a_then_b}
\end{figure}

\section{Case Studies} \label{sec:examples}

In this section, we demonstrate the simulated results of the examples presented in Sec.~\ref{sec:problem_formulation} using the UAV. 
These examples illustrate how an autonomous agent with a complex task could operate over an uncertain environment. We also show that our algorithm is efficient for an online synthesis.
In both scenarios, we have $N = 10$ and~$|\mathbf{U}| = 3$ (i.e., straight, left, and right). 
We show the average clock time for synthesis using a standard desktop with $3.4$ GHz Intel CPU and 16 GB RAM.

\subsection{Surveillance} 
The simulated result for the surveillance mission (Example~\ref{exa:surveillance}) is shown in Fig.~\ref{fig:exec_surv} where the targets are shown in red, blue, and magenta. At each time~$t$, the candidate trajectories are shown in black, the trajectory with maximum relaxed satisfaction probability is shown in bold red, the executed trajectory (i.e., past run) is shown in bold green, and the camera's field of view is shown in yellow. The true locations of the targets are marked in bold. The initial state of the UAV is~$[30m, 10m, 70 \deg]^T$. We assume that the targets are randomly moving at a known maximum velocity. Based on the velocity, the belief changes over time.

The initial control of the UAV is synthesised based on poorly estimated beliefs of the targets shown in Fig.~\ref{fig:exec_surv_1}. 
The beliefs are updated as the UAV makes observations, and the UAV always uses the most up-to-date beliefs to synthesise a new control input at any given time. Figure~\ref{fig:exec_surv} shows how the beliefs change over time as observations are made. It also shows how the changing beliefs affect the way the trajectories are generated.

The average synthesis time at each~$t$ is $2.94s$ where the minimum and maximum are~$1.22s$ and~$6.67s$, respectively.
For synthesising the initial control input at time~$t=1$, the number of candidate trajectories generated is~$683$, which took~$4.85s$ to compute. However, if we had to solve Problem~\ref{problem:finite} without limiting the number of candidate trajectories as shown in (\ref{eqn:prune}), the required number of candidate trajectories in synthesis would have been~$3^{70}$ ($\approx 2.5 \times 10^{33}$) which is not scalable in practice. 
To demonstrate the improvement in synthesis time, we ran the simulation to compute the average synthesis time without the limiting constant~$N$. However, no solution was given in a practical time (still running after 5 hours). 
Although the synthesis without the limiting constant~$N$ would provide a complete and optimal solution, the approach is not scalable for an online synthesis.

\subsection{Prioritised search}
We demonstrate the simulated result for Example~\ref{exa:search} in Fig.~\ref{fig:exec_a_then_b}. The blue and red contours represent the beliefs over true locations of Tom and Jerry, respectively. Like in the previous example, we assume that Tom and Jerry are moving at a known maximum walking velocity. The initial state of the UAV is the same as the previous example.

Initially, the UAV assigned a task to locate Tom whose true location is poorly estimated. Based on the belief over Tom's true location, the UAV heads straight to the region where the probability of finding Tom is the highest (i.e., around $[75, 70]$). At~$t = 10$, Tom is in the view of the UAV, but the UAV fails to detect him due to sensor noise. With the series of observations, the belief over Tom's location is updated (compare Fig.~\ref{fig:exec_a_then_b_1} with~\ref{fig:exec_a_then_b_19}), and the more accurate belief is used to synthesise a better control input.
At~$t=29$, the UAV finally finds Tom, and then it generates a new control input to find Jerry while updating the beliefs. Note that the belief over Jerry's location at time~$t=29$ is wider than that at time~$t=1$ because Jerry's walking speed is reflected in estimating Jerry's belief. At time~$t=37$, Jerry is found, and the mission ends.
The average synthesis time at each time~$t$ is $6.25s$ where the minimum and maximum are~$1.95s$ and~$34.74s$, respectively.

\section{CONCLUSIONS} \label{sec:conclusion}

In this paper, we proposed a probabilistic extension to signal temporal logic to specify complex tasks over target beliefs. We also presented an efficient 
receding horizon synthesis algorithm that maximises the probability of satisfying a specification in this logic. 
Through simulations using a simple UAV model, we showed that the algorithm can easily adapt to changes in the belief space.
This work is an important step towards synthesis of complex tasks over a belief space, but many open problems remain. In the future, we will consider cases where the conditional independence assumption is violated. We will consider models that include uncertain external disturbaces such as wind.


\bibliographystyle{IEEEtran}
{\small \bibliography{Chanyeol_reference}}

\end{document}